\documentclass[12pt,a4paper]{article}

\usepackage{amssymb}
\usepackage{amsmath}
\usepackage[nosort]{cite}
\usepackage[hyperref,bulletsep]{collect}
\def\gfxon{\usepackage[final]{graphicx}}

\gfxon

\makeatletter
\let\old@startsection=\@startsection
\renewcommand{\@startsection}[6]{\old@startsection{#1}{#2}{#3}{#4}{#5}{#6\mathversion{bold}}}
\makeatother

\makeatletter \@addtoreset{equation}{section} \makeatother

\allowdisplaybreaks[3]

\makeatletter
\let\old@makecaption=\@makecaption
\def\@makecaption{\small\old@makecaption}
\makeatother

\let\oldPhi=\Phi
\let\oldPsi=\Psi
\let\oldGamma=\Gamma
\let\oldDelta=\Delta
\let\oldSigma=\Sigma
\let\oldLambda=\Lambda
\let\oldTheta=\Theta
\let\oldPi=\Pi
\renewcommand{\Phi}{\mathnormal{\oldPhi}}
\renewcommand{\Psi}{\mathnormal{\oldPsi}}
\renewcommand{\Gamma}{\mathnormal{\oldGamma}}
\renewcommand{\Sigma}{\mathnormal{\oldSigma}}
\renewcommand{\Delta}{\mathnormal{\oldDelta}}
\renewcommand{\Theta}{\mathnormal{\oldTheta}}
\renewcommand{\Lambda}{\mathnormal{\oldLambda}}
\renewcommand{\Pi}{\mathnormal{\oldPi}}

\newcommand{\ham}{\mathcal{H}}
\newcommand{\fldZ}{\mathcal{Z}}

\newcommand{\superN}{\mathcal{N}}

\newcommand{\order}[1]{\mathcal{O}(#1)}
\newcommand{\Integers}{\mathbb{Z}}

\newcommand{\gammafn}{\oldGamma}

\ifx\genfrac\sdflkaj
\newcommand{\atopfrac}[2]{{{#1}\above0pt{#2}}}
\else
\newcommand{\atopfrac}[2]{\genfrac{}{}{0pt}{}{#1}{#2}}
\fi
\newcommand{\sfrac}[2]{{\textstyle\frac{#1}{#2}}}

\newcommand{\indup}[1]{_{\mathrm{#1}}}
\newcommand{\indups}[1]{_{\mathrm{\scriptscriptstyle #1}}}

\newcommand{\MMM}[2]{{\arraycolsep0pt\begin{array}[b]{c}\makebox[0cm]{$\atopfrac{#2}{\downarrow}$}\\#1\end{array}}}

\newcommand{\lrbrk}[1]{\left(#1\right)}
\newcommand{\bigbrk}[1]{\bigl(#1\bigr)}
\newcommand{\Bigbrk}[1]{\Bigl(#1\Bigr)}

\newcommand{\PTerm}[1]{\{#1\}}

\newcommand{\state}[1]{\mathopen{|}#1\mathclose{\rangle}}

\newcommand{\alg}[1]{\mathfrak{#1}}
\newcommand{\grp}[1]{\mathrm{#1}}

\newcommand{\nn}{\nonumber}
\newcommand{\nln}{\nonumber\\}
\newcommand{\nl}[1][0pt]{\nonumber\\[#1]&\hspace{-4\arraycolsep}&\mathord{}}
\newcommand{\nlnum}{\\&\hspace{-4\arraycolsep}&\mathord{}}
\newcommand{\earel}[1]{\mathrel{}&\hspace{-2\arraycolsep}#1\hspace{-2\arraycolsep}&\mathrel{}}
\newcommand{\eq}{\earel{=}}
\newcommand{\beq}{\begin{equation}}
\newcommand{\eeq}{\end{equation}}

\def\[{\begin{equation}}
\def\]{\end{equation}}
\def\<{\begin{eqnarray}}
\def\>{\end{eqnarray}}

\newenvironment{bulletlist}{\begin{list}{$\bullet$}{\leftmargin1.5em\itemsep0pt}}{\end{list}}

\makeatletter
\def\mr@ignsp#1 {\ifx\:#1\@empty\else #1\expandafter\mr@ignsp\fi}%
\newcommand{\multiref}[1]{\begingroup
\xdef\mr@no@sparg{\expandafter\mr@ignsp#1 \: }%
\def\mr@comma{}%
\@for\mr@refs:=\mr@no@sparg\do{\mr@comma\def\mr@comma{,}\ref{\mr@refs}}%
\endgroup}
\makeatother

\newcommand{\hypref}[2]{\ifx\href\asklfhas #2\else\href{#1}{#2}\fi}

\newcommand{\secref}[1]{\S\multiref{#1}}

\newcommand{\appref}[1]{App.~\S\multiref{#1}}

\newcommand{\tabref}[1]{Tab.~\multiref{#1}}

\newcommand{\figref}[1]{Fig.~\multiref{#1}}
\ifx\eqref\sdflkaj
\newcommand{\eqref}[1]{(\multiref{#1})}
\else
\renewcommand{\eqref}[1]{(\multiref{#1})}
\fi

\ifx\href\asklfhas\newcommand{\href}[2]{#2}\fi
\newcommand{\arxivno}[1]{\href{http://arxiv.org/abs/#1}{#1}}

\def\dalf{\textstyle{\frac{d}{2}}}

\def\pic#1#2{\hbox{\lower#1pt\hbox{~\mbox{\epsfxsize=20truemm \epsffile{#2}}}}}
\def\pic#1#2#3{\hbox{\lower#1pt\hbox{~\mbox{\includegraphics[scale=#3]{#2}}}}}

\begin{document}

\thispagestyle{empty}
\begin{flushright}\footnotesize
\texttt{\arxivno{arxiv:0705.0321}}\\
\texttt{AEI-2007-026}\\
\texttt{PUTP-2233}%
\vspace{0.5cm}
\end{flushright}

\renewcommand{\thefootnote}{\fnsymbol{footnote}}
\setcounter{footnote}{0}

\begin{center}
{\Large\textbf{\mathversion{bold}
The Four-Loop Dressing Phase of $\superN=4$ SYM%
}\par}

\vspace{1.5cm}

\textsc{N.~Beisert$^{1,2}$, T.~McLoughlin$^{3}$ and R.~Roiban$^{3}$} \vspace{8mm}

\textit{$^{1}$ Max-Planck-Institut f\"ur Gravitationsphysik\\%
Albert-Einstein-Institut\\%
Am M\"uhlenberg 1, 14476 Potsdam, Germany}%
\vspace{3mm}%

\textit{$^{2}$ Joseph Henry Laboratories\\%
Princeton University\\%
Princeton, NJ 08544, USA}%
\vspace{3mm}

\textit{$^{3}$ Department of Physics\\
The Pennsylvania State University\\ 
University Park, PA 16802, USA}%
\vspace{8mm}

\texttt{nbeisert@aei.mpg.de}\\
\texttt{tmclough@phys.psu.edu}\\
\texttt{radu@phys.psu.edu}%
\vspace{20mm}


\textbf{Abstract} \vspace{5mm}

\begin{minipage}{12.7cm}
We compute the dilatation generator in the 
$\alg{su}(2)$ sector of planar $\superN=4$ super Yang-Mills theory
at four-loops.
We use the known world-sheet scattering matrix 
to constrain the structure of the generator. 
The remaining few coefficients can be computed 
directly from Feynman diagrams.
This allows us to confirm  
previous conjectures for the leading contribution 
to the dressing phase which is proportional to $\zeta(3)$.
\end{minipage}

\end{center}

\vspace{0.5cm}

\newpage
\setcounter{page}{1}
\renewcommand{\thefootnote}{\arabic{footnote}}
\setcounter{footnote}{0}

\tableofcontents

\section{Introduction and Overview}

The means available for analyzing the AdS/CFT correspondence 
improved dramatically with the discovery of perturbative integrability of
the gauge theory dilatation operator 
\cite{Minahan:2002ve,Beisert:2003yb,Beisert:2003tq}
and that of classical
integrability of the world sheet sigma model
\cite{Mandal:2002fs,Bena:2003wd}. 
Furthermore, there are arguments \cite{Berkovits:2004xu} on the
string theory side of the correspondence that a infinite family of 
BRST invariant, non-local  currents exists at all orders in the inverse 
't Hooft coupling expansion
suggesting that  integrability persists in the quantum theory. 
In the absence of a definitive and constructive 
proof of all-order integrability, one may nonetheless assume it and
study its consequences. 

The fundamental quantity in an integrable (discrete or continuous)
theory defined on infinitely extended space-like slices is the
scattering matrix of excitations. The S-matrix is constrained by the
symmetries of the theory; integrability further requires that no
particle production occurs in the scattering process and that the
$n\rightarrow n$ scattering process is realized by repeated
$2\rightarrow 2$ scattering events. 
A necessary requirement is that the two-particle S-matrix obeys the Yang-Baxter equation. 

For the AdS/CFT correspondence the relevant two-particle scattering
matrix was introduced in \cite{Staudacher:2004tk}; it turns out that
the global symmetries -- a centrally extended form of
$\alg{psu}(2|2)^2$ determine it up to an overall phase
\cite{Beisert:2005tm}.
The Yang-Baxter equation holds automatically in this case. Although
initially the S-matrix was determined in the gauge theory framework it
was later shown that the tensor structure agreed with perturbative
calculations in the gauge-fixed world-sheet theory \cite{Klose:2006zd}
and that it is consistent with the Zamolodchikov-Faddeev algebra for
the string sigma-model \cite{Arutyunov:2006yd}.


In relativistic quantum field theories the analogous ``dressing factor''
is determined by crossing symmetry, information on the spectrum of
bound states and perhaps perturbative calculations. For the AdS/CFT
correspondence both the world sheet and gauge theory integrable
systems do not exhibit Lorentz invariance. While on the gauge theory
side there is little reason to require an analog of crossing symmetry,
on the string theory side two-dimensional Lorentz invariance is only
spontaneously broken. As such, one may expect that some form of
crossing symmetry survives this breaking. 

A crossing-like equation
was constructed in \cite{Janik:2006dc} and 
shown in \cite{Arutyunov:2006iu}
to hold for the known leading
\cite{Arutyunov:2004vx}
and next-to-leading terms \cite{Hernandez:2006tk,Gromov:2007cd}.
An all-orders solution at a strong coupling expansion 
was proposed in \cite{Beisert:2006ib}.

An unfortunate feature of this solution is that it is an asymptotic
series and thus, without additional information, cannot be directly
used to define the dressing phase everywhere in the coupling constant
space. In \cite{Beisert:2006ez} an analytic continuation scheme was
described which allowed a guess for the weak-coupling expansion of the
dressing phase whose contribution to anomalous dimensions starts at
four-loop order where it predicts a transcendental contribution
proportional to $\zeta(3)$. This prediction remarkably agrees with
the direct calculation of the four-loop cusp anomalous dimension
\cite{Bern:2006ew, Cachazo:2006az}. 
Subsequently the expansions at weak and strong coupling were shown to
be fully consistent \cite{Kotikov:2006ts}
and an integral expression for the phase at finite coupling
was proposed in \cite{Belitsky:2007zp,Dorey:2007xn}

In fact the above agreement is slightly surprising:
The four-gluon scattering amplitude of \cite{Bern:2006ew} 
is related to the infinite-spin limit of twist-two anomalous dimensions.
Conversely, the analysis of \cite{Beisert:2006ez}
strictly applies to local operators of twist three or higher.
Due to the asymptotic nature of the higher-loop Bethe equations
the twist-two anomalous dimension can only be predicted reliably up
to three loops, see \cite{Kotikov:2007cy} for further recent developments.
The agreement thus implies that the cusp anomalous dimension
is universal for operators of all twists. In other words,
the limiting procedure described in \cite{Eden:2006rx,Beisert:2006ez}
does not suffer from potential order-of-limits ambiguities.

As remarkable as it is, this agreement also presents a puzzle: The
universality of the dressing phase implies that all anomalous
dimensions of $\mathcal{N}=4$ SYM have, at four-loop order, a
transcendental contribution proportional to $\zeta(3)$. While this is
not at all surprising for non-compact subsectors of $\mathcal{N}=4$ SYM
in the large spin limit, it does seem surprising for finite spins and
for compact sectors. 
Indeed, in the infinite spin limit the RG flow
mixes an infinite number of operators allowing transcendental numbers
to appear even if they are absent at the level of the anomalous
dimension matrix.  
In the latter cases however, the RG flow mixes only
finitely many operators and thus precludes the appearance of
transcendental numbers. Consequently, for the conjectured dressing phase to be
correct, $\zeta(3)$ must appear at the level of the anomalous
dimension matrix elements.

Loop integrals may be interpreted -- in a first quantized language --
as a sum over infinitely many intermediate states producing an analogy
with the large-spin $\alg{sl}(2)$ sector operator mixing. 
From this standpoint, one is entitled
to expect the appearance of transcendental numbers at some
sufficiently high loop order in any sector. 
One of the building blocks of the
calculation of the renormalization factors of scalar composite
operators is the one-loop scalar bubble diagram. It turns out that, in
dimensional regularization, 
its $\epsilon$-expansion contains $\zeta(3)$ at $\order{\epsilon^2}$;
consequently, if this bubble is part of a larger diagram and the other
momentum integrals yield a third-order pole in the $\epsilon$
expansion, $\zeta(3)$ may appear in the residue of a first order pole
and thus may contribute to some entry of the anomalous dimension
matrix. Counting the required number of inverse powers of the
dimensional regulator we immediately reach the conclusion that 
this mechanism may function first at the four-loop order. 

In this paper we shall compute the four-loop dilatation operator in
the $\alg{su}(2)$ sector and show that the expectations outlined above are
indeed realized. We shall begin in \secref{sec:ham} with a review of the
constraints imposed by $\alg{su}(2)$ symmetry and Feynman
diagrammatics. The unknown coefficients are parametrized in terms of
the first nontrivial coefficient of the dressing phase. 
However, unlike earlier discussions \cite{Beisert:2004jw,Beisert:2005wv} we
shall not assume that this operator is part of an integrable
Hamiltonian. Instead, we shall determine in \secref{sec:loop} the
unknown coefficients 
-- and in particular the coefficient related to the dressing phase -- by
a direct calculation. The calculation is dramatically simplified by
the observation that the unknown coefficients may be associated to
so-called maximal interactions (i.e.~interactions that reshuffle the
spins in a maximal way). \secref{sec:concl} contains our conclusions. Some technical
details as well as some momentum integrals useful for going beyond
four loops are included in the appendices.

\section{Long-Range Heisenberg Hamiltonian}
\label{sec:ham}

A full-fledged field theory calculation of the complete four-loop
planar dilatation generator in $\superN=4$ SYM is a difficult task
whose completion clearly requires new, deep insight in higher-loop technology.
The main complications are the extensive combinatorics and 
the intricate algebra of loop momenta
inherent to gauge theories at higher perturbative orders.  However,
our primary goal is to compute the relevant coefficient for the
dressing at this order.  The dressing factor can be observed in all
closed sectors of the model and we can conveniently restrict to the
simplest one, the $\alg{su}(2)$ subsector, cf.~\cite{Beisert:2004ry}.
It consists of local operators which are made from just two complex
scalars, let us denote them by $\fldZ$ and $\phi$, or, equivalently,
\emph{spin up} and \emph{spin down}.  Here the planar dilatation
operator turns into the Heisenberg XXX$_{1/2}$ Hamiltonian
\cite{Minahan:2002ve} with perturbative long-ranged deformations
\cite{Beisert:2003tq}
\[\label{eq:PertHam}
\ham=\sum_{\ell=0}^\infty \lrbrk{\frac{\lambda}{16\pi^2}}^\ell\ham_\ell.
\]
Determining this Hamiltonian at the fourth
perturbative order would provide us with the leading piece of the
dressing phase.

The first few perturbative deformations of the Hamiltonian were
obtained in \cite{Beisert:2003tq}: This construction made use of the
fact that the Hamiltonian is some linear combination of all
interactions compatible with $\alg{su}(2)$ symmetry which can
originate from Feynman diagrams. The coefficients of the interactions
could in principle be computed from perturbative field
theory. However, such an elaborate calculation was avoided by matching
the coefficients to make the spectrum of the Hamiltonian agree with
some available data. Together with the further assumption of
integrability, a proposal for the Hamiltonian at the third
perturbative order could be made.  The conjecture has since passed
various tests
\cite{Beisert:2003ys,Eden:2004ua,Kotikov:2004er,Bern:2005iz,Eden:2006rx} 
which prove that it is correct. 

Here we shall repeat the above procedure to constrain the fourth-order
Hamiltonian as much as possible without making any unproven assumptions. 
The crucial new input that allows us to go to higher orders 
is the picture of asymptotic excitation states \cite{Staudacher:2004tk}
and its scattering matrix \cite{Beisert:2005tm}.
In this picture, spin chain states are replaced by excitations 
above a ferromagnetic vacuum, the magnons. The ferromagnetic vacuum consists of a long
chain of aligned spins, say $\fldZ$
\[
\state{0}=\state{\ldots\fldZ\fldZ\fldZ\ldots}.
\]
This state is protected by a half-BPS condition 
from receiving quantum corrections to its energy;
the complete cancellation of corrections 
to two-point functions in field theory at two loops is 
demonstrated explicitly in \cite{D'Hoker:1998tz,Penati:1999ba,Penati:2000zv}.
A single-magnon state has one of these spins flipped to $\phi$,
say at position $k$
\[
\state{k}=\state{\ldots\fldZ\MMM{\phi}{k}\fldZ\ldots}.
\]
Similarly, one can construct states with two or more magnons
\[
\state{k,\ell,\ldots}=\state{\ldots\fldZ\MMM{\phi}{k}\fldZ\ldots\fldZ\MMM{\phi}{\ell}\fldZ\ldots\fldZ\MMM{\phi}{\ldots}\fldZ\ldots\ldots\ldots}.
\]
In the asymptotic coordinate space Bethe ansatz \cite{Staudacher:2004tk}
the magnons are arranged into momentum eigenstates 
with an additional phase shift when two magnons move
past each other. 

The excitation picture is highly constrained by its residual symmetry.
It was shown in \cite{Beisert:2005tm} that 
the form of the one and two-magnon states
is almost completely determined. 
The only degrees of freedom are a finite redefinition of the
coupling constant and the dressing phase. 
The possibility to redefine coupling constants 
by a finite amount is inherent to field theories.
We can make a suitable choice and
all other choices can be recovered from it by substitution.
A general analysis \cite{Beisert:2005wv}
shows that the dressing phase starts to contribute 
at four loops with a single undetermined coefficient $\beta_{2,3}$.

Note that these results are actually not based on the (unproven) 
assumption of higher-loop integrability: 
integrability or factorized scattering constrains the
scattering of three or more particles. It also implies
a constraint on the two-particle scattering matrix
which, however, in this case is automatically satisfied
\cite{Beisert:2005tm}.

\begin{table}\centering
\<
\ham_0\eq +\PTerm{}
\nn\\[5pt]
\ham_1\eq +2\PTerm{}-2\PTerm{1}
\nn\\[5pt]
\ham_2\eq 
-8\PTerm{}
+12\PTerm{1}
-2\bigbrk{\PTerm{1,2}+\PTerm{2,1}}
\nn\\[5pt]
\ham_3\eq 
+60\PTerm{}
-104\PTerm{1}
+4\PTerm{1,3}
+24\bigbrk{\PTerm{1,2}+\PTerm{2,1}}
\nl
-4i\epsilon_2\PTerm{1,3,2}
+4i\epsilon_2\PTerm{2,1,3}
-4\bigbrk{\PTerm{1,2,3}+\PTerm{3,2,1}}
\nn\\[5pt]
\ham_4\eq
+\bigbrk{-560-4\beta_{2,3}} \PTerm{} 
\nl
+\bigbrk{+1072+12\beta_{2,3}+8\epsilon\indup{3a}} \PTerm{1} 
\nl
+\bigbrk{-84-6\beta_{2,3}-4\epsilon\indup{3a}} \PTerm{1,3} 
\nl
-4\PTerm{1,4}
\nl
+\bigbrk{-302-4\beta_{2,3}-8\epsilon\indup{3a}} \bigbrk{ \PTerm{1,2} + \PTerm{2,1} }  
\nl
+\bigbrk{+4\beta_{2,3}+4\epsilon\indup{3a}+2i\epsilon\indup{3c}-4i\epsilon\indup{3d}} \PTerm{1,3,2} 
\nl
+\bigbrk{+4\beta_{2,3}+4\epsilon\indup{3a}-2i\epsilon\indup{3c}+4i\epsilon\indup{3d}} \PTerm{2,1,3}
\nl
+\bigbrk{4-2i\epsilon\indup{3c}} \bigbrk{ \PTerm{1,2,4} + \PTerm{1,4,3} }  
\nl
+\bigbrk{4+2i\epsilon\indup{3c}} \bigbrk{ \PTerm{1,3,4} + \PTerm{2,1,4} }  
\nl
+\bigbrk{+96+4\epsilon\indup{3a}} \bigbrk{ \PTerm{1,2,3} + \PTerm{3,2,1} }  
\nl
+\bigbrk{-12-2\beta_{2,3}-4\epsilon\indup{3a}} \PTerm{2,1,3,2} 
\nl
+\bigbrk{+18+4\epsilon\indup{3a}} \bigbrk{ \PTerm{1,3,2,4} + \PTerm{2,1,4,3} }  
\nl
+\bigbrk{-8 - 2\epsilon\indup{3a}-2i\epsilon\indup{3b}} \bigbrk{ \PTerm{1,2,4,3} + \PTerm{1,4,3,2} }  
\nl
+\bigbrk{-8 - 2\epsilon\indup{3a}+2i\epsilon\indup{3b}} \bigbrk{ \PTerm{2,1,3,4} + \PTerm{3,2,1,4} }  
\nl
-10\bigbrk{ \PTerm{1,2,3,4} + \PTerm{4,3,2,1} }
\nn
\>

\caption{The four-loop Hamiltonian.
The coefficient $\beta_{2,3}$ is the leading
coefficient of the dressing phase at weak coupling. 
We confirm the prediction $\beta_{2,3}=4\zeta(3)$ \protect\cite{Beisert:2006ez} 
as the principal result of this paper.
The coefficients $\epsilon$ correspond to similarity transformations
and do not influence the spectrum.}
\label{tab:Ham}
\end{table}

We can now match the coefficients of the Hamiltonian to the
zero-, one- and two-particle states.
The analysis proceeds along the lines of \cite{Beisert:2005wv}
and the most general result is shown in \tabref{tab:Ham}.
The interaction symbols $\PTerm{a,b,c,\ldots}$
represent a sequence of nearest-neighbor interactions
$\mathcal{P}_{p}$ of spins at sites $p$ and $p+1$
summed homogeneously over the spin chain of length $L$
\[
\PTerm{a,b,c,\ldots}=
\sum_{p=1}^L
\mathcal{P}_{p+a}
\mathcal{P}_{p+b}
\mathcal{P}_{p+c}
\cdots\,.
\]
As undetermined parameters it contains 
the coefficient $\beta_{2,3}$ for the dressing phase
as well as several irrelevant parameters $\epsilon$. 
The latter correspond to similarity transformations of the Hamiltonian 
which do not affect its spectrum. One may change their 
values by applying the similarity transformation 
$\ham\mapsto\exp(-i\mathcal{X})\ham\exp(+i\mathcal{X})$
with the second and third-order contributions to $\mathcal{X}$
given by
\<
\mathcal{X}_2\eq
\delta\epsilon\indup{2}\bigbrk{\PTerm{1,2}+\PTerm{2,1}},
\nln
\mathcal{X}_3\eq
i\delta\epsilon\indup{3a}\bigbrk{\PTerm{2,1,3}-\PTerm{1,3,2}}
+\delta\epsilon\indup{3b}\bigbrk{\PTerm{1,2,3}+\PTerm{3,2,1}}
\nl
+\delta\epsilon\indup{3c}\PTerm{1,3}
+\delta\epsilon\indup{3d}\bigbrk{\PTerm{1,2}+\PTerm{2,1}}.
\>

It is worth pointing out that the Hamiltonian 
at fourth order can be fixed uniquely up to irrelevant terms. 
In other words, the scattering of three or more magnons
is fixed by the scattering of two magnons. 
This feature is related to the $\alg{su}(2)$ symmetry of the
interactions: Interactions at four loops act on at most five adjacent spins. 
Any elementary interaction among three or more magnons 
(and therefore at most two vacuum spins)
is related to an interaction among at most two magnons
(and at least three vacuum spins)
by flipping all five interacting spins.
Starting at five loops this picture breaks down because
interactions of six spins allow for elementary interactions of three magnons 
which leave no trace on the sector with two or fewer magnons.
It turns out that our four-loop Hamiltonian in \tabref{tab:Ham} 
is integrable, i.e.~it is of the form determined 
(but not displayed explicitly) in \cite{Beisert:2005wv}.
We have therefore proved four-loop integrability 
in the $\alg{su}(2)$ sector.

\begin{figure}\centering
\parbox{2cm}{\centering\includegraphics{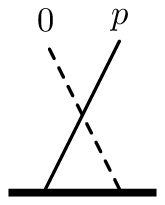}\\\textbf{1}: $\PTerm{1}$}
\parbox{2.75cm}{\centering\includegraphics{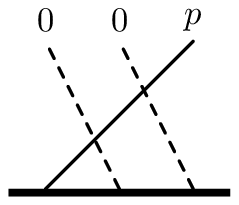}\\\textbf{2}: $\PTerm{2,1}$}
\parbox{3.5cm}{\centering\includegraphics{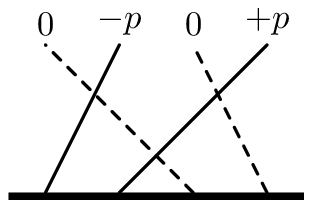}\\\textbf{3a}: $\PTerm{1,3,2}$}
\parbox{3.5cm}{\centering\includegraphics{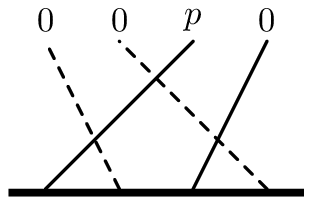}\\\textbf{3b}: $\PTerm{2,1,3}$}
\parbox{3.5cm}{\centering\includegraphics{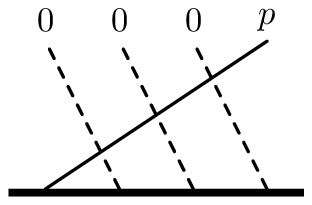}\\\textbf{3c}: $\PTerm{3,2,1}$}
\vspace{0.5cm}

\parbox{3.5cm}{\centering\includegraphics{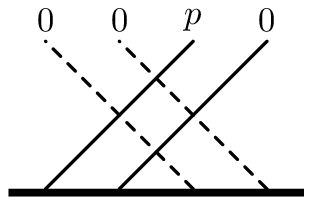}\\\textbf{4a}: $\PTerm{2,1,3,2}$}
\parbox{4.25cm}{\centering\includegraphics{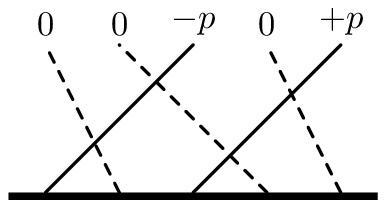}\\\textbf{4b}: $\PTerm{2,1,4,3}$}
\parbox{4.25cm}{\centering\includegraphics{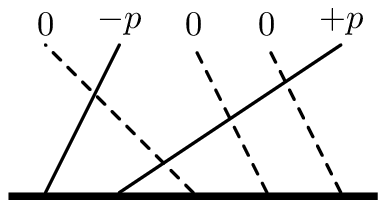}\\\textbf{4c}: $\PTerm{1,4,3,2}$}
\vspace{0.5cm}

\parbox{4.25cm}{\centering\includegraphics{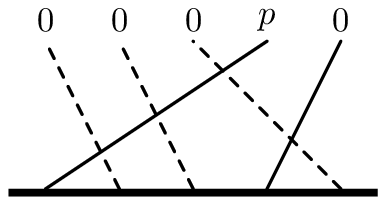}\\\textbf{4d}: $\PTerm{3,2,1,4}$}
\parbox{4.25cm}{\centering\includegraphics{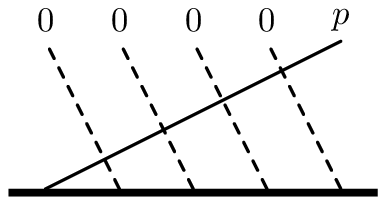}\\\textbf{4e}: $\PTerm{4,3,2,1}$}

\caption{Maximal planar interactions up to four loops.
Below the diagrams the permutation symbols are indicated.
Solid and dashed lines correspond to two complex scalars
in $\superN=4$ SYM. 
Above the diagrams suitable momenta to remove IR singularities are indicated.
}
\label{fig:maxi}
\end{figure}

The four-loop Hamiltonian in \tabref{tab:Ham} is fixed to a large
extent.  To determine the dressing phase coefficient $\beta_{2,3}$ it
suffices to compute only a small number of its coefficients. We see
that $\beta_{2,3}$ couples, among others, to the very first and fifth
but last interaction structure in \tabref{tab:Ham}.  The first
structure does not redistribute the spins along the spin chain. There
are exceedingly many planar Feynman diagrams which do not change
flavor, for example those containing only interactions of gluons and
scalars.  Therefore, a direct computation of this coefficient seems
particularly difficult.
In contrast, the coefficients of the five last interactions
can be computed relatively easily. They form a class of interactions
which reshuffle the spins in a maximal way.  At $\ell$ loops, they
contain $\ell$ permutations of nearest neighbors, see
\figref{fig:maxi} for a graphical representation of their induced
permutations.  This is the maximum reshuffling allowed by planar
Feynman diagrams
\cite{Beisert:2003tq} and it will turn out to
be \emph{generated by the quartic interactions of the scalars only}.
In other words, the relevant Feynman diagrams will
be those of a $\phi^4$-theory.

Moreover, the individual maximal interactions are identifiable
by acting on special states: 
Assume that the Hamiltonian density maps a state 
\[
\ham_4\state{\ldots\phi\phi\fldZ\fldZ\ldots}=c\state{\ldots\fldZ\fldZ\phi\phi\ldots}+\ldots \,.
\]
There is a single interaction which achieves this particular reshuffling
of spins: $\PTerm{2,1,3,2}$,
cf.~\figref{fig:maxi}. Therefore we could infer that $c$ 
equals the coefficient of this interaction,
$c=-12-2\beta_{2,3}-4\epsilon\indup{3a}$.
The same is true for the other maximal interactions:
If all lines going right are associated with $\phi$ and the 
others with $\fldZ$ then a $\phi$ will move past a $\fldZ$ 
towards the right at each elementary crossing. 
The effect will thus uniquely identify the corresponding interaction.

Being the representation of the dilatation generator on the gauge
invariant operators in the 
$\alg{su}(2)$ sector, the spin chain Hamiltonian in
\tabref{tab:Ham} is also the anomalous dimension matrix of operators in
this sector. In any conformal field theory the eigenvalues of the
anomalous dimension matrix are independent of the renormalization
scheme. Its matrix elements however do not generically have this
property. At the level of the Hamiltonian in \tabref{tab:Ham} this is
reflected by the fact that the undetermined coefficients $\epsilon$
do not affect its eigenvalues \cite{Beisert:2004jw}.

\section{Four-Loop Calculation}
\label{sec:loop}

With these preparations we are now in a position to compute the
undetermined coefficients that appear in the spin chain Hamiltonian.
Among the various approaches to finding the anomalous dimension matrix
we shall consider the renormalization of composite operators. 
It was successfully used in \cite{Gross:2002su} 
to determine the two-loop and 
(under the assumption of proper BMN scaling for one-excitation BMN states) 
the all-loop dispersion relation. 

The renormalization of composite operators and the subtraction of
subdivergences proceeds by introducing renormalization factors and
counterterm diagrams analogous to the Bogoliubov R-operation. 
For our purpose this procedure was systematized in \cite{Beisert:2003tq} 
where an iterative subtraction scheme 
was developed that allows the subtraction of entire subdiagrams. 
This is the scheme we shall use.

We are therefore to compute Feynman diagrams with one vertex
being the composite operator of interest and 
additional vertices dictated by the $\mathcal{N}=4$ SYM Lagrangian.
As described in detail in the previous section, our goal is to find 
the entries of the four-loop anomalous dimension matrix that reshuffle scalar
fields in a maximal way.  Besides scalar fields, the internal lines
of these diagrams may \emph{a priori} also be fermions and gauge fields.
Two simple observations imply however 
that the situation is substantially simpler.

By inspection of the diagrams in \figref{fig:maxi} it is 
easy to see that Feynman diagrams containing gauge fields cannot
lead to such maximal reshuffling (in the sense described previously) 
of scalar fields. Indeed, using the fact that the gauge field
interactions are flavor-blind, one may see that replacing any of the 
four-point vertices  by scalar-vector interactions leads to diagrams
not exhibiting maximal reshuffling.

Let us consider next scalar-fermion interactions. R-charge
conservation implies that any diagram with external fermion fields
and an insertion of an operator in the $\alg{su}(2)$ sector vanishes
identically. Diagrams with internal fermion lines have a similar fate.
To see this let us note that the Yukawa interactions of the 
$\mathcal{N}=4$ SYM Lagrangian are proportional to the $\grp{SO}(6)$ Dirac matrices. 
Since the fields of the $\alg{su}(2)$ sector are complex, 
the Dirac matrices appearing in these vertices will also carry complex vector indices.
Their algebra, $\{\Gamma^{\bar a},\,\Gamma^{b}\}=\eta^{\bar a b}$,
implies that holomorphic matrices square to zero.
Therefore the flavor of scalar fields coupling
to fermionic loops must alternate. It is then easy to see that
for $2n$ flavors at most $n$ pairs
can be interchanged. 
This does not lead to a maximal permutation and we can thus 
disregard fermion loops.

The conclusion is therefore that the Feynman diagrams contributing to
the entries of the four-loop anomalous dimension matrix describing a maximal
reshuffling of spins are scalar diagrams in which each vertex contains
two types of scalar fields and the interaction interchanges
them. These are the diagrams listed in \figref{fig:maxi}.

To compute the contribution to the anomalous dimension matrix we need
to compute the amplitudes in  \figref{fig:maxi} and isolate the
overall ultraviolet divergence by subtracting all their
UV subdivergences. While in general it is convenient to use a variant
of dimensional regularization which preserves supersymmetry, in the
context of our calculation making a definite choice is not an
issue since all our diagrams have scalar internal lines.
However, since all fields are massless
we must be careful to separate the UV divergences from IR
divergences. To this end we shall assign off-shell momenta to
some of the external fields;%
\footnote{The external fields can 
either belong to the operator being renormalized or be attached to 
the vertices of the $\mathcal{N}=4$ SYM Lagrangian.}
they are chosen such that the number of
momenta is minimal while all IR divergences are eliminated. It turns
out that up to four-loop order it suffices that only two of the
external fields carry momentum; an appropriate 
choice is depicted in \figref{fig:maxi}.  

All momentum integrals may be computed easily by reduction to  a small
set of master integrals. Common building blocks are bubble diagrams with
arbitrary exponents for the two propagators; their expressions are 
(here and elsewhere the dimensionality of space-time is $d=4-2\epsilon$)
\footnote{We extract  the overall momentum dependence for the
sake of notational convenience.}
\<
 L(a_1,a_2) 
\earel{\equiv} (p^2){}^{a_{12}-d/2}\times \hbox{\pic{14}{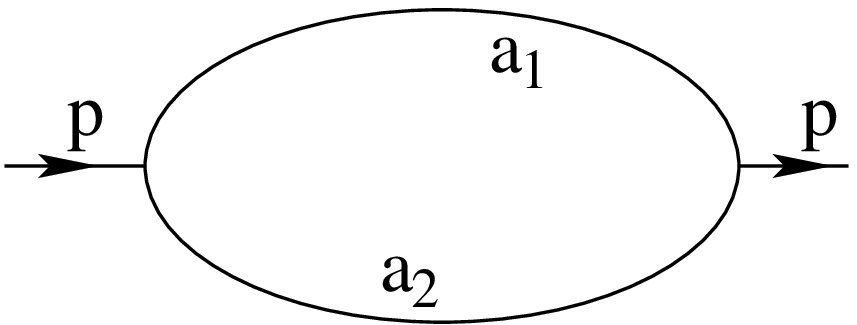}{.38}}
\nln
\eq
\int\frac{d^d q}{(2\pi)^d}\, \frac{(p^2){}^{a_{12}-d/2}}{(q^2)^{a_1}((q+p)^2)^{a_2}}
\\
\eq
(4\pi)^{-d/2}\,
\frac{\gammafn(a_{12}-\dalf)}{\gammafn(a_{1})\gammafn(a_{2})}\,
\frac{\gammafn(\dalf-a_{1})\gammafn(\dalf-a_{2})}{\gammafn(d-a_{12})}
\nonumber
\>
where $a_{12}=a_1+a_2$. 
Indeed, diagrams \textbf{1}, \textbf{2}, 
\textbf{3b}, \textbf{3c}, \textbf{4d} and \textbf{4e}
may be computed exactly by repeated identification of one-loop 
bubble subdiagrams. Once a bubble subintegral is evaluated, the
exponent of the propagator carrying the momentum $p$ flowing through the
bubble is shifted by an integer
multiple of the dimensional regulator.

A similar iterative identification of bubble subintegrals reduces the diagrams 
\textbf{3a} and \textbf{4c} to special cases of the two-loop master bubble integral
\[
T(a_1,a_2,a_3, a_4,a_5)\;\equiv  (p^2){}^{a_{12345}-d}\times \hbox{\pic{14}{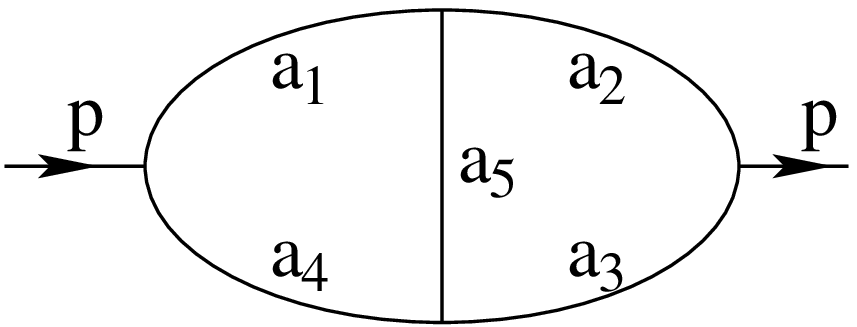}{.38}}
\label{Tint}
\]
with some arbitrary powers of propagators. For integer exponents
$a_{1,\dots ,5}$ such integrals have been computed in the 
past (e.g. \cite{Smirnov:2004ym, Smirnov:2006ry}). 
We are however interested in situations when some of the exponents
are not integers (special cases have been previously analyzed 
in \cite{Kazakov:1983pk, Kotikov:1995cw}), 
being dependent on the dimensional regulator.
Perhaps the
most effective way of computing such integrals is to use the
Mellin-Barnes parametrization \cite{Smirnov:2004ym, Smirnov:2006ry}. 
The identity
\[
\frac{1}{(a+b)^\nu}=\frac{1}{2\pi i}\int_{-i\infty}^{+i\infty}dw\,
\frac{\gammafn(-w)\gammafn(w+\nu)}{\gammafn(\nu)}\,
\frac{a^w}{b^{\nu+w}}
\]
allows a straightforward 
evaluation of the Feynman parameter integrals and expresses the result
of the momentum integral in terms of multiple contour integrals which
can be evaluated through the residue theorem. This method has the
advantage of producing explicit integral representations for the 
coefficients of the various powers of the dimensional regulator. 
The algorithm of \cite{Tausk:1999vh} for the analytic continuation $\epsilon\rightarrow 0$ 
as well as the numerical evaluation of the resulting coeffcients 
has been successfully automated \cite{Anastasiou:2005cb,Czakon:2005rk}.
An MB parametrization of $T(a_1,a_2,a_3, a_4,a_5)$ is
\begin{eqnarray}
\earel{}T(a_1,a_2,a_3,a_4,a_5)=\frac{(4\pi)^{-d}}
{\gammafn(a_1) \gammafn(a_4)\gammafn(a_5)\gammafn(d-a_{145})}
\\
\earel{}\qquad\times\int_{-i\infty}^{+i\infty}\frac{dw_1dw_2}{(2\pi i)^2}
\frac{\gammafn(a_{145} - \dalf + w_{12})}
{\gammafn(a_{1245} - \dalf + w_{12})}
\gammafn(-w_1)\gammafn(-w_2)
\gammafn(a_4 + w_{12})\gammafn(d- a_{1245}- w_{12})
\nln
\earel{}\qquad~~~~~~~~~~~~~~~~~~
\frac{\gammafn(\dalf- a_{14} - w_1)\gammafn(a_{12345} - d + w_1)}
{\gammafn(\textstyle{\frac{3d}{2}}- a_{12345} - w_1) }
\frac{\gammafn(\dalf - a_{45} - w_2)\gammafn(\dalf-a_3 + w_2)}
{\gammafn(a_3 - w_2)}
\nonumber
\end{eqnarray}
with the notation $a_{ijk...}=a_i+a_j+a_k+...$ and similarly for $w_{ijk...}$. 
While this parametrization does not manifestly exhibit the
symmetries of the original diagram, they are restored after the
remaining integrals are performed. It is possible (though perhaps less
efficient in terms of the necessary number of MB parameters) to construct 
a Mellin-Barnes parametrization manifestly
exhibiting the $(\Integers_2)^2$ symmetries of \eqref{Tint}.
The $\epsilon$-expansions of the two-loop integrals we shall require 
read:
\<
T(1,1,1,1,\epsilon)\eq
L(1,1)^2\,\Bigbrk{
\sfrac{1}{3}+\sfrac{1}{3}\epsilon+\sfrac{1}{3}\epsilon^2
+\bigbrk{-\sfrac{7}{3}+\sfrac{14}{3}\zeta(3)}\epsilon^3+\ldots},
\nln
T(1,1,1,\epsilon,1)\eq L(1,1)^2\,\Bigbrk{
\sfrac{1}{6}+\sfrac{1}{2}\epsilon+\sfrac{13}{6}\epsilon^2
+\bigbrk{+\sfrac{55}{6}-\sfrac{23}{3}\zeta(3)}\epsilon^3+\ldots},
\nln
T(1,1,1,1,2\epsilon)\eq L(1,1)^2\,\Bigbrk{
\sfrac{1}{6}+\sfrac{1}{3}\epsilon+\sfrac{1}{3}\epsilon^2
+\bigbrk{-\sfrac{17}{3}+\sfrac{31}{3}\zeta(3)}\epsilon^3+\ldots},
\nln
T(1,1,1,1+\epsilon,\epsilon)\eq L(1,1)^2\,\Bigbrk{
\sfrac{5}{24}+\sfrac{5}{12}\epsilon+\sfrac{25}{24}\epsilon^2
+\bigbrk{+\sfrac{5}{12}+\sfrac{19}{6}\zeta(3)}\epsilon^3+\ldots},
\nln
T(1,1,1,2\epsilon,1)\eq L(1,1)^2\,\Bigbrk{
\sfrac{1}{12}+\sfrac{5}{12}\epsilon+\sfrac{29}{12}\epsilon^2
+\bigbrk{+\sfrac{161}{12}-\sfrac{71}{6}\zeta(3)}\epsilon^3+\ldots}.
\>
Here it was convenient to factor out two powers of the one-loop 
bubble $L(1,1)$ which has the expansion 
\<
L(1,1)\eq
(4\pi)^{\epsilon}\,
\frac{\gammafn(\epsilon)\gammafn^2(1-\epsilon)}{16\pi^2\gammafn(2-2\epsilon)}
\\\nn
\eq
\frac{1}{16\pi^2\epsilon}
\lrbrk{4\pi e^{-\gamma}}^\epsilon
\Bigbrk{1+2\epsilon+\bigbrk{4-\sfrac{1}{12}\pi^2}\epsilon^2
 +\bigbrk{8-\sfrac{1}{6}\pi^2-\sfrac{7}{3}\zeta(3)}\epsilon^3+\ldots}.
\>

In both of the last two integrals 
\textbf{4a} and \textbf{4b}
it is trivial to isolate  a factor  $L(1,1)$. 
The remaining three-loop integrals may be computed in
several ways. 
One approach makes use of integration by parts
identities, known in this case as the triangle rule, 
to reduce them to combinations of one- and two-loop bubble integrals
with various exponents (see \appref{app:triangle} 
for details). 
A second approach directly evaluates the three-loop integrals 
and in the process tests that the infrared region is
non-singular. 
We list the necessary Mellin-Barnes integrals 
and the $\epsilon$-expansions of all diagrams
in \appref{app:VariousIntegrals}. 
Needless to say, the two 
calculations lead to the same answer.

For a vector of operators $\mathbf{O}$, the relation between the bare and
renormalized operators is given by the renormalization factor $\mathbf{Z}$
\[
\mathbf{O}^{\mathrm{bare}}=\mathbf{Z}\cdot \mathbf{O}^{\mathrm{ren}}~.
\]
The $\ell$-loop contribution to $\mathbf{Z}$ is found from the overall
divergence of $\ell$-loop diagrams with exactly one insertion of a member
of the vector $\mathbf{O}$. To isolate the overall divergence it is necessary
to include counterterm diagrams which are generated recursively 
by the lower-loop renormalization factor. These diagrams also
eliminate the non-local momentum dependence. The relation between the
renormalization factor and the anomalous dimension matrix 
(a.k.a.~dilatation generator or spin chain Hamiltonian) 
$\ham$
is standard:
\[
\delta\ham=\lim_{\epsilon\rightarrow 0}
\epsilon\, \mathbf{Z}^{-1}\frac{d}{d\ln g\indups{YM}}\,\mathbf{Z}~.
\label{anomdimmat}
\]
This expression implies an exponential-like structure for the
renormalization factor ${\bf Z}$;%
\footnote{For operators which do not mix under RG flow the relation is
$Z=\exp\big[{\epsilon^{-1}
\int_0^1 dt\,t^{-1}\gamma(tg\indups{YM})}\big]$.} 
in particular, the
derivative of ${\bf Z}$ must be left-proportional to ${\bf Z}$ and, in order that the
$\delta\ham$ be well defined in the $\epsilon\rightarrow 0$ limit, the factor of
proportionality can only have additional
simple poles.%
\footnote{It is in principle 
possible that in a different renormalization scheme individual 
matrix elements could have divergent terms; however these terms should 
be removable by similarity transformations.}

A subtraction scheme that enforces these constraints and at each loop
order isolates directly the contribution to the anomalous dimension matrix
was described in \cite{Beisert:2003tq} for the use in two-point
functions. An adapted version for the use in operator renormalization diagrams
is presented in \appref{app:tetris} where the explicit rules
are given and then applied to the relevant diagrams.

Finally, in this scheme the equation \eqref{anomdimmat} reduces to the
simple operation of picking the residue of the $1/\epsilon$ pole of
the subtracted diagrams
\[
\tilde I=
2(16\pi^2)^{\ell}
\lim_{\epsilon\to 0}\epsilon\,\bar I(\epsilon)~.
\]
The factor of $(16\pi^2)^\ell$ corresponds to the normalization
of the $\ell$-loop Hamiltonian $\ham_\ell$ 
and allows for a direct comparison of the quantity $\tilde I$ 
to the coefficients in \eqref{eq:PertHam}.
In our case this leads to
\<
\begin{array}[b]{rclcrcl}
\tilde I_1\eq-2,&& \tilde I\indup{4a}\eq -\phantom{0}4+\phantom{0}4\,\zeta(3),\\[2pt]
\tilde I_2\eq-2,&& \tilde I\indup{4b}\eq +10-12\,\zeta(3),\\[2pt]
\tilde I\indup{3a}\eq+4,&& \tilde I\indup{4c}\eq+\phantom{0}2+\phantom{0}8\,\zeta(3),\\[2pt]
\tilde I\indup{3b}\eq-4,&& \tilde I\indup{4d}\eq-10+\phantom{0}4\,\zeta(3),\\[2pt]
\tilde I\indup{3c}\eq-4,&& \tilde I\indup{4e}\eq-10,
\end{array}
\>
which represent the coefficients relating the structures listed in
\figref{fig:maxi} and the spin chain Hamiltonian (cf.~\tabref{tab:Ham}). 
Clearly, $\tilde I_1$, $\tilde I_2$, $\tilde I\indup{3c}$ 
and $\tilde I\indup{4e}$ reproduce the coefficients of
$\PTerm{1}$, $\PTerm{2,1}$, $\PTerm{3,2,1}$ and $\PTerm{4, 3,2,1}$,
respectively. The coefficients undetermined by symmetry considerations
are fixed by our calculation to be%
\footnote{The conventional factors of $i$ indicate 
that the Hamiltonian is not manifestly hermitian.
With a proper choice of scalar product, however, it becomes quasi-hermitian
as it should.}
\[
i\epsilon_{2}=-1,~~~~
\epsilon_{3a}=-2-3\zeta(3),~~~~
i\epsilon_{3b}=-3-\zeta(3)
\label{parameters}
\]
and
\[
\beta_{2,3}=4\zeta(3).
\label{parameterbeta}
\]
In particular we are able to uniquely fix the leading coefficient $\beta_{2,3}$ for the
dressing phase. It is in full agreement with
the results of \cite{Bern:2006ew,Beisert:2006ez,Cachazo:2006az}.

\section{Conclusions and Outlook}
\label{sec:concl}

In this paper we have computed the four-loop dilatation operator in 
the $\alg{su}(2)$ sector of $\mathcal{N}=4$ SYM. The main observation 
which led to substantial technical simplifications is that the 
coefficients undetermined by symmetry constraints can be chosen to
correspond to ``maximal interactions'' -- i.e. interactions
that reshuffle the spins in a maximal way. For appropriately chosen
gauge theory operators these interactions are entirely determined by Feynman
diagrams with only scalar interactions.
We found that, starting at four-loop order, the anomalous dimensions 
of long operators become transcendental; this may be traced to the
dilatation operator acquiring 
transcendental coefficients. We have
extracted the relevant coefficient of the dressing phase and
found it identical to the one reproducing the four-loop cusp
anomalous dimension computed in \cite{Bern:2006ew, Cachazo:2006az}. 
Our result confirms the particular analytic continuation used to
guess the dressing phase at weak coupling \cite{Beisert:2006ez}.

The main obstacle for computing higher-loop anomalous dimensions in
any sector of $\mathcal{N}=4$ SYM and thus directly computing the
S-matrix dressing phase is, as in all off-shell calculations, the
proliferation of Feynman diagrams. In compact sectors the symmetries of
the theory restrict (sometimes substantially) 
the structure of the anomalous dimension matrix. At any loop order the
maximal interactions enjoy the same technical simplifications as the
ones employed in the calculations described here; moreover, it is
possible that some of the relevant momentum integrals 
exhibit a recursive structure.%
\footnote{A candidate for this property is 
$\PTerm{m,\dots, 1,m+1,\dots,2,\dots,n+m-1,\dots,n-1, n+m\dots n}$.}
It would be interesting to identify and
compute them, thus providing a direct evaluation 
of important parts of the dressing phase.

\paragraph{Acknowledgments:}

N.B.~would like to thank DESY Zeuthen for the hospitality during the
CAPP 2007 school.
N.B.~acknowledges support by a Sloan Research Fellowship
while at Princeton University.
The work of R.R.~is supported in part by the U.S.~National Science
Foundation Grant No.~PHY-0608114. Any opinions,
findings and conclusions or recommendations expressed in this
material are those of the authors and do not necessarily reflect the
views of the National Science Foundation.

\appendix

\section{The Triangle Rule}
\label{app:triangle}

Consider the Feynman integral 
\[
F(a_1,a_2, a_3,a_4,a_5)\equiv \hbox{\pic{31}{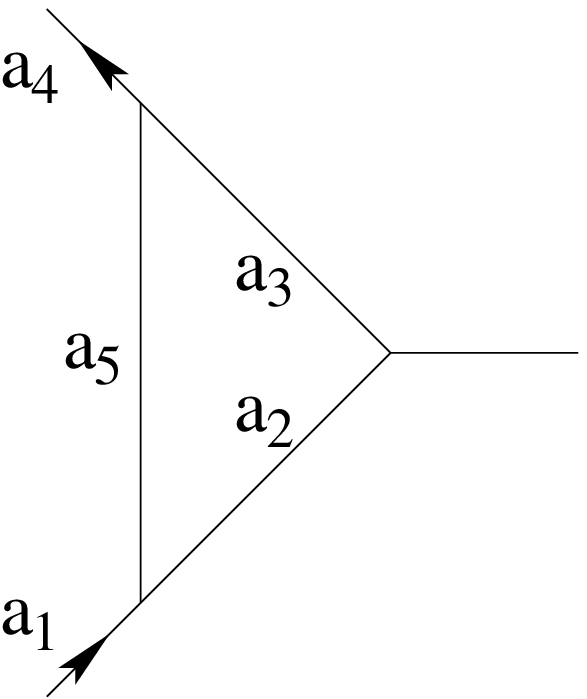}{.35}}~~.
\]
This integral may be part of a larger Feynman diagram and the labels
$a_{1,\dots,5}$ represent the exponents of the propagators of the
corresponding internal lines.   
Inserting the operator $l^\mu\partial/\partial l^\mu$ in the
integral representing this diagram and equating the results of
the action of the derivative on the original integrand and the result of
the integration by parts leads to
\[
(d-a_2-a_3-2a_5)F=
\left(a_2\ \mathbf{2^+}\left(\mathbf{5^-}-\mathbf{1^-}\right)+a_3\ \mathbf{3^+}\left(\mathbf{5^-}-\mathbf{4^-}\right)\right)F
\]
  where, for example,
$\mathbf{1^\pm}F(a_1,a_2,a_3,a_4,a_5)=F(a_1\pm1,a_2,a_3,a_4,a_5)$.

The various terms in such a decomposition may
acquire however spurious infrared divergences 
which are regularized by the dimensional
regulator and -- provided that the IR of the original
integral was properly regularized -- cancel when all terms are
assembled.

The triangle rule together with the straightforward evaluation of bubble
integrals leads to the following expressions 
for the diagrams in \figref{fig:maxi}:%
\footnote{It is trivial to identify in the
expressions of $I\indup{4a}$ and $I\indup{4c}$ the IR divergent
components mentioned above.}
\<\label{eq:trisimp}
I_1\eq (p^2){}^{-\epsilon}\,L(1,1),
\nln
I_2\eq (p^2){}^{-2\epsilon}\,L(1,1)\,L(1+\epsilon,1),
\nln
I\indup{3a}\eq (p^2){}^{-3\epsilon}\,L(1,1)\,T(1,1,1,1,\epsilon),
\nln
I\indup{3b}\eq (p^2){}^{-3\epsilon}\,L(1,1)^2L(1+\epsilon,1+\epsilon),
\nln
I\indup{3c}\eq (p^2){}^{-3\epsilon}\,L(1,1)\,L(1+\epsilon,1)\,L(1+2\epsilon,1),
\nln
I\indup{4a}\eq (p^2){}^{-4\epsilon}\,\frac{L(1,1)}{1-3\epsilon}
\Big[  
L(2,\epsilon)\,T(1,1,1,2\epsilon,1)
-L(2,3\epsilon)\,T(1,1,1,\epsilon,1)
\Big],
\nln
I\indup{4b}\eq (p^2){}^{-4\epsilon}\,L(1,1)^2T(1,1,1,1+\epsilon,\epsilon),
\nln
I\indup{4c}\eq (p^2){}^{-4\epsilon}\,\frac{L(1,1)}{1-3\epsilon}
\Big[ L(2,\epsilon)\,T(1,1,1,1,2\epsilon)
  -L(2,1)\,T(1,1,1,1+\epsilon,\epsilon)
\nl~~~~~~~~~~~~ 
+\epsilon\, L(1,1+\epsilon)\,T(1,1,1,1,2\epsilon)
  -\epsilon\, L(1,1+\epsilon)\,T(1,1,1,2\epsilon,1)\Big],
\nln
I\indup{4d}\eq (p^2){}^{-4\epsilon}\,L(1,1)^2L(1+\epsilon,1)\,L(1+2\epsilon,1+\epsilon),
\nln
I\indup{4e}\eq (p^2){}^{-4\epsilon}\,L(1,1)\,L(1+\epsilon,1)\,L(1+2\epsilon,1)\,L(1+3\epsilon,1).
\>

Note that the integrals 
$T(1,1,1,\epsilon,1)$ and $T(1,1,1,2\epsilon,1)$ can be evaluated
further using the triangle rule
\<
T(1,1,1,\epsilon,1)\eq
\frac{L(1,1)}{1-3\epsilon}
\Bigbrk{
L(2,\epsilon)-L(2,2\epsilon)
+\epsilon\, L(1,1+\epsilon)-\epsilon\, L(1+\epsilon,1+\epsilon)},
\\\nn
T(1,1,1,2\epsilon,1)\eq
\frac{L(1,1)}{1-4\epsilon}
\Bigbrk{
L(2,2\epsilon)-L(2,3\epsilon)
+2\epsilon\, L(1,1+2\epsilon)-2\epsilon\, L(1+\epsilon,1+2\epsilon)}.
\>
For compactness, we left them unexpanded in \eqref{eq:trisimp};
The integral $I\indup{4a}$ can thus be evaluated as 
an analytic expression in $\epsilon$.

\section{Three-Loop Integrals}
\label{app:VariousIntegrals}

The three-loop integral that remains after one identifies a one-loop
bubble subintegral in $I\indup{4a}$ may be evaluated directly thus
testing the application of the triangle rule and the correct infrared
regularization of the contributions to the anomalous dimension
matrix. A Mellin-Barnes parametrization of a master
integral containing $I\indup{4a}$ is
\begin{eqnarray}
BM(a_1,a_2, a_3) \earel{\equiv} (p^2){}^{a_{123}+4-3d/2}\times \hbox{\pic{20}{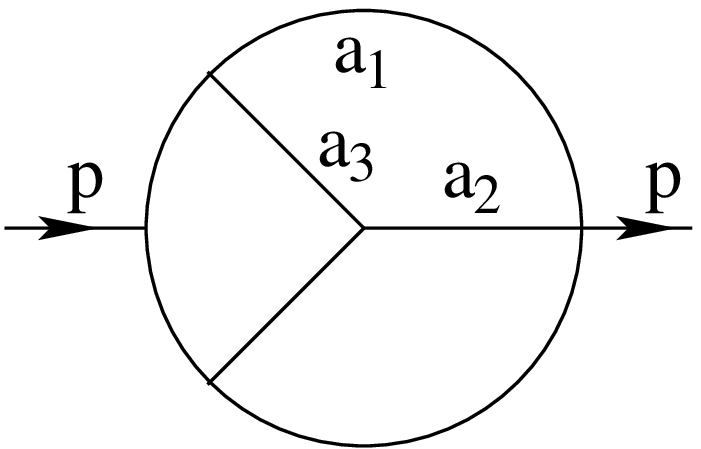}{.38}}
\nln
\eq(4\pi)^{-3d/2}~
\int_{-i\infty}^{i\infty}\frac{dw_1dw_2dw_3dw_4}{(2\pi i)^4}
\nl\times
\frac{\gammafn(-w_1)\gammafn(-w_2)\gammafn(-w_3)\gammafn(-w_4)
\gammafn( a_{123} - \dalf + w_{12})}
{\gammafn(a_1) \gammafn(a_2) \gammafn(a_3) \gammafn(d-a_{123} )}
\nl\times
\frac{
\gammafn(\dalf-a_{13} - w_1)
\gammafn(\dalf-a_{12} - w_2)
\gammafn(a_1 + w_{12})}{\gammafn(1 - w_1)
\gammafn(3\dalf-3 - a_{123} - w_2)
\gammafn(1 + a_{123} - \dalf + w_{12})}
\nl\times
\frac{\gammafn(d-2 - a_{123} - w_{123})
\gammafn( \dalf-1 +  w_3)}{\gammafn(1 - w_3)}
\nl\times
\frac{\gammafn(d-2 - a_{123} - w_{24})
\gammafn(\dalf + w_{24})}{\gammafn(-w_{24})}
\nlnum\nn\times
\frac{\gammafn(3 + a_{123} - d + w_{234})
\gammafn(1 + a_{123} - \dalf + w_{1234} )
\gammafn(1-\dalf - w_{234})}
{\gammafn(1 - w_3)
\gammafn(d-1 + w_{234})}~.
\end{eqnarray}
Then,
\[
I\indup{4a}= (p^2){}^{-4\epsilon}\,L(1,1)BM(1,\epsilon, 1) 
\]
and its evaluation leads to the result listed in equation \eqref{expanded}.

Similarly, $I\indup{4c}$ is a special case of the master integral 
\[
BL(a_1,a_2,
a_3,a_4,a_5)\;\equiv (p^2){}^{a_{12345}+3-3d/2}\times\hbox{\pic{15}{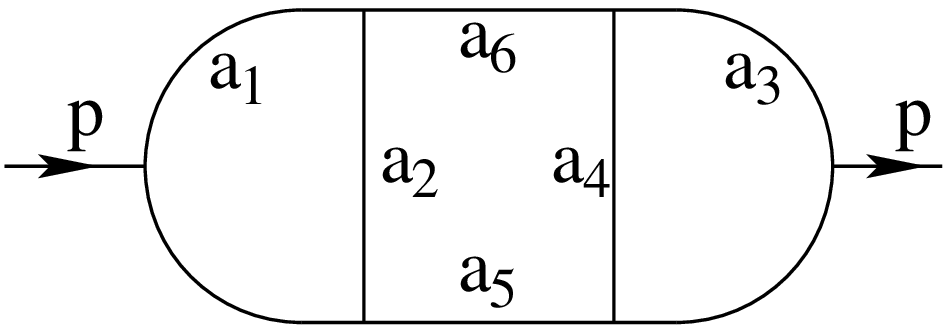}{.38}}~~.
\]
A Mellin-Barnes parametrization is
\begin{eqnarray}
&&BL(a_1,a_2, a_3,a_4,a_5)=(4\pi)^{-3d/2}~
\int_{-i\infty}^{i\infty}\frac{dw_1dw_2dw_3dw_4}{(2\pi i)^4}
\nln
&\times&\frac{\gammafn(-w_1)\gammafn(-w_2)\gammafn(1 - \dalf + a_{12}+ w_{12} )
\gammafn(1 + w_{12})}{\gammafn(a_1)\gammafn(a_2)\gammafn(d-1 - a_{12} )}
\nln
&\times&\frac{
\gammafn(-w_3)\gammafn(-w_4)\gammafn(1 - \dalf + a_{34}+ w_{34} )
\gammafn(1 + w_{34})}{\gammafn(a_3)\gammafn(a_4)\gammafn(d-1 - a_{34} )}
\nln
&\times&
\gammafn( \dalf-1 - a_1 - w_1)
\gammafn( \dalf-1 - a_2 - w_2)
\gammafn(\dalf-1 - a_3 - w_3)
\gammafn(\dalf-1 - a_4 - w_4)
\nln
&\times&
\frac{\gammafn(2 + a_{123456} - \textstyle{\frac{3d}{2}} + w_{13})}
{\gammafn(2 d-2 - a_{123456}  - w_{13})}
\frac{\gammafn(\dalf -a_5 + w_{24})}{\gammafn(a_5 - w_{24})}
\frac{\gammafn( \textstyle{\frac{3d}{2}} -  2-a_{12346} - w_{1234})}
{\gammafn(2 + a_{12346} - d + w_{1234})}~~.
\end{eqnarray}
Then
\[
I\indup{4c}=(p^2){}^{-4\epsilon}\,L(1,1)\lim_{\nu\rightarrow 0} 
BL(1,\epsilon, 1,1,\nu, 1)
\]
whose evaluation leads to the result listed in equation \eqref{expanded}.

The integrals listed here are useful for the calculation of the
coefficients of the higher-loop Hamiltonian in the $\alg{su}(2)$ sector.
The resulting $\epsilon$-expansions of the integrals 
in \figref{fig:maxi} read
\<
I_1\eq 
\frac{1}{16\pi^2\epsilon}
\lrbrk{\frac{4\pi e^{-\gamma}}{p^2}}^\epsilon
\Bigbrk{1+2\epsilon+\bigbrk{4-\sfrac{1}{12}\pi^2}\epsilon^2
 +\bigbrk{8-\sfrac{1}{6}\pi^2-\sfrac{7}{3}\zeta(3)}\epsilon^3+\ldots}
,
\nln
I_2\eq
(I_1)^2\Bigbrk{
\sfrac{1}{2}+\sfrac{1}{2}\epsilon+\sfrac{3}{2}\epsilon^2
+\bigbrk{\sfrac{9}{2}-3\zeta(3)}\epsilon^3+\ldots},
\nln
I\indup{3a}\eq
(I_1)^3\Bigbrk{
\sfrac{1}{3}+\sfrac{1}{3}\epsilon+\sfrac{1}{3}\epsilon^2
+\bigbrk{-\sfrac{7}{3}+\sfrac{14}{3}\zeta(3)}\epsilon^3+\ldots},
\nln
I\indup{3b}\eq
(I_1)^3\Bigbrk{
\sfrac{1}{3}+\sfrac{2}{3}\epsilon+\sfrac{8}{3}\epsilon^2
+\bigbrk{\sfrac{32}{3}-\sfrac{22}{3}\zeta(3)}\epsilon^3+\ldots},
\nln
I\indup{3c}\eq
(I_1)^3\Bigbrk{
\sfrac{1}{6}+\sfrac{1}{2}\epsilon+\sfrac{13}{6}\epsilon^2
+\bigbrk{\sfrac{55}{6}-\sfrac{11}{3}\zeta(3)}\epsilon^3+\ldots},
\nln
I\indup{4a}\eq
(I_1)^4\Bigbrk{
\sfrac{1}{12}+\sfrac{1}{3}\epsilon+\sfrac{19}{12}\epsilon^2
+\bigbrk{\sfrac{43}{6}-\sfrac{10}{3}\zeta(3)}\epsilon^3+\ldots},
\nln
I\indup{4b}\eq
(I_1)^4\Bigbrk{
\sfrac{5}{24}+\sfrac{5}{12}\epsilon+\sfrac{25}{24}\epsilon^2
+\bigbrk{\sfrac{5}{12}+\sfrac{19}{6}\zeta(3)}\epsilon^3+\ldots},
\nln
I\indup{4c}\eq
(I_1)^4\Bigbrk{
\sfrac{1}{8}+\sfrac{1}{3}\epsilon+\sfrac{9}{8}\epsilon^2
+\bigbrk{\sfrac{10}{3}-\sfrac{3}{2}\zeta(3)}\epsilon^3+\ldots},
\nln
I\indup{4d}\eq
(I_1)^4\Bigbrk{
\sfrac{1}{8}+\sfrac{1}{2}\epsilon+\sfrac{21}{8}\epsilon^2
+\bigbrk{\sfrac{27}{2}-\sfrac{13}{2}\zeta(3)}\epsilon^3+\ldots},
\nln
I\indup{4e}\eq
(I_1)^4\Bigbrk{
\sfrac{1}{24}+\sfrac{1}{4}\epsilon+\sfrac{37}{24}\epsilon^2
+\bigbrk{\sfrac{107}{12}-\sfrac{13}{6}\zeta(3)}\epsilon^3+\ldots}.
\label{expanded}
\>

\section{Subtraction Scheme}
\label{app:tetris}

Here we describe the subtraction scheme used to 
extract the contributions to the anomalous dimensions
without having to insert couterterms at each stage of the calculation.
For each connected diagram drawn with the composite
operator as the lower-most vertex, one
\begin{bulletlist}
\item partitions it in all possible connected subdiagrams 
(including the trivial partition into a single subdiagram)
and interprets those diagrams as contributing 
to the renormalization of a composite operator,

\item discards all partitions which are interconnected horizontally
(all partial diagrams must be ``dropped'' onto the composite operator
from above in a well-defined sequence, in similarity to a famous arcade game),

\item discards all partitions for which there are two or more
top-most diagrams,

\item evaluates the momentum integrals for the of the remaining partitions,

\item sums the products of the momentum integrals for each
partition weighted by $\ell(-1)^n$ 
($n$ is the number of partial diagrams and
$\ell$ is the loop number of the top-most diagram in the partition).%
\footnote{It is easy to see that this weight is related to a derivative with
respect to the loop-counting parameter, as in equation \protect\eqref{anomdimmat}.}

\end{bulletlist}

Applying this scheme to the diagrams in \figref{fig:maxi} 
we find the following subtracted integrals ${\tilde I}$ 
\<
\bar I_1\eq -I_1 ,
\nln
\bar I_2\eq -2I_2+I_1^2,
\nln
\bar I\indup{3a}\eq -3I\indup{3a}+2I_2I_1,
\nln
\bar I\indup{3b}\eq -3I\indup{3b}+4I_2I_1-I_1^3,
\nln
\bar I\indup{3c}\eq -3I\indup{3c}+3I_2I_1-I_1^3,
\nln
\bar I\indup{4a}\eq -4I\indup{4a}+I\indup{3a}I_1+3I\indup{3b}I_1+4I\indup{2}^2-6I_2I_1^2+I_1^4,
\nln
\bar I\indup{4b}\eq -4I\indup{4b}+3I\indup{3a}I_1+I\indup{3b}I_1+2I_2^2-2I_2I_1^2,
\nln
\bar I\indup{4c}\eq -4I\indup{4c}+I\indup{3a}I_1+I\indup{3c}I_1+2I\indup{2}^2-I_2I_1^2,
\nln
\bar I\indup{4d}\eq -4I\indup{4d}+3I\indup{3b}I_1+3I\indup{3c}I_1+2I\indup{2}^2-5I_2I_1^2+I_1^4,
\nln
\bar I\indup{4e}\eq -4I\indup{4e}+4I\indup{3c}I_1+2I\indup{2}^2-4I_2I_1^2+I_1^4.
\>
It is not hard to find that the the quantities ${\tilde I}$ 
exhibit only a simple pole in the $\epsilon$-expansion.
A strong crosscheck of the correctness of the subtraction 
is the cancellation of non-local and divergent momentum-dependence 
that arises in \eqref{expanded}.

\bibliography{4loopzeta}
\bibliographystyle{nb}

\end{document}